\documentclass[letterpaper]{article}

\usepackage{natbib,alifeconf}  %% The order is important
\usepackage [autostyle]{csquotes}

\title{Competitive Exclusion in an Artificial Foraging Ecosystem }
\author{John C. Stevenson \\ Long Beach Institute \\jcs@alumni.caltech.edu}
\begin{document}
\maketitle

\begin{abstract}
	Artificial ecosystems provide an additional experimental tool to support laboratory work, field work, and theoretical development in competitive exclusion research. A novel application of a spatiotemporal agent based model is presented which simulates two foraging species of different intrinsic growth rates competing for the same replenishing resource in stable and seasonal environments. This experimental approach provides precise control over the parameters for the environment, the species, and individual movements. Detailed trajectories of these non-equilibrium populations and their characteristics are produced. Narrow zones of potential coexistence are identified within the environmental and intrinsic growth parameter spaces. An example of commensalism driven by the local spatial dynamics is identified. Results of these experiments are discussed in context of modern coexistence theory and research in movement-mediated community assembly. Constraints on possible origination scenarios are identified. 
	
%\keywords{competitive exclusion, spatiotemporal niches}
	
\end{abstract}
\section{Introduction }

Artificial ecosystems bring precise and controllable experiments to support laboratory and field work \citep{siep,grainger,godwin2020empiricist,schlagel} and theoretical modeling in competitive exclusion and coexistence \citep{chessMech,chesson2018}, including temporal and spatial variations and stochastic effects \citep{hening2021general,chesson2000general,ellner2022toward}. Reported here are empirical results of persistence and extinction events obtained from novel, discrete, spatiotemporal, stochastic simulations of a two-species, finite populations in constant and seasonal environments. These simulations use a unique underlying mechanism with individual movement decisions that generate dynamics that are consistent with standard, mathematical, non-equilibrium models of biology and ecology; and highlight the role simple artificial life simulations can play in complex ecological research.

The theory of competitive exclusion \citep{gause,hardin,volt1} and its apparent absence in the natural world has been a subject of debate and research since the publication of the theory of natural selection. Darwin himself was troubled by the implications species diversity had on his theory. Hutchinson's observations on plankton and suggestions that its species diversity was driven by the lack of equilibrium brought on by seasons \citep{hutch,armstrong} have been pursued well into the twenty-first century \citep{mayfield,bohn}. This potential lack of equilibrium suggests generative modeling as an approach \citep{epstein}. 

A leading framework for theory on competitive exclusion in community ecology is \textquote{modern coexistence theory} (MCT) \citep{chessMech,chesson2018,barab}. This framework formulates the Lokta-Volterra equations for population density, using a concept of resident and invasive species to define stable coexistence \citep{armstrong,kan}, and maps these requirements onto the coefficients (or functions) in Lokta-Volterra equations. This approach is often applied to temporal variation of the environment and less frequently to spatial variations \citep{chesson1981models,chesson2000general}. Numerical solutions of the Lokta-Volterra equations with (assumed) models and distributions provide a second approach for understanding competitive exclusion \citep{ovask,huisman}. For two species consuming the same resource over multiple seasons, direct numerical integration of the Lokta-Volterra equations with constant coefficients produced one of the first explorations of coexistence for varying environments over time \citep{koch}.  The species densities were numerically solved over the spring/summer growing season, then these densities were reduced 40-fold for the start of next season. This process was repeated for many seasons. Stable cycles, particularity with noise added to the spring initial populations, enabled identification of narrow zones of coexistence based on Lokta-Volterra coefficients.

Recognition of the importance of organismal movement for understanding community assembly, species coexistence, and biodiversity has increased the need for modeling of local movement such that it can be scaled up to the metacommunity level \citep{schlagel,chesson1981models,chesson2005scale}. Recent simulations using a lottery model demonstrated the importance of spatial effects though the lottery model used is intended for very large populations and does not track individuals \citep{ellner2022toward}. Of the various approaches to model individual movement \citep{patterson,luo,grimm}, the approach used here is a minimal model of a two-species foraging ecosystem \citep{roughgarden} which conforms to the movement ecology paradigm \citep{nathan}. This approach results in emergent behavior which overcomes the \textquote{lack (of) mechanistic descriptions of competition parameters} in MCT \citep{schlagel} and allows investigation of species coexistence for discrete, scholastic models of finite populations. Models with individual movement dynamics within finite populations, however, require a modified definitions in order to accommodate the inevitable extinction of a finite stochastic population over a \textquote{very long time} \citep{fisher1923,wright1931,cole}.

The objectives of this research are to demonstrate emergent competitive exclusion and coexistence behaviors in a specific and novel simulation based on a finite population with individual movement and seasonal variations using an agent based model (ABM), to place these behaviors within Modern Control Theory and movement mediated community assembly contexts, and to identify the non-equilibrium dynamics of these finite stochastically-driven populations. 
\begin{figure}
	\begin{center}
		\includegraphics[angle=0,scale=0.4]{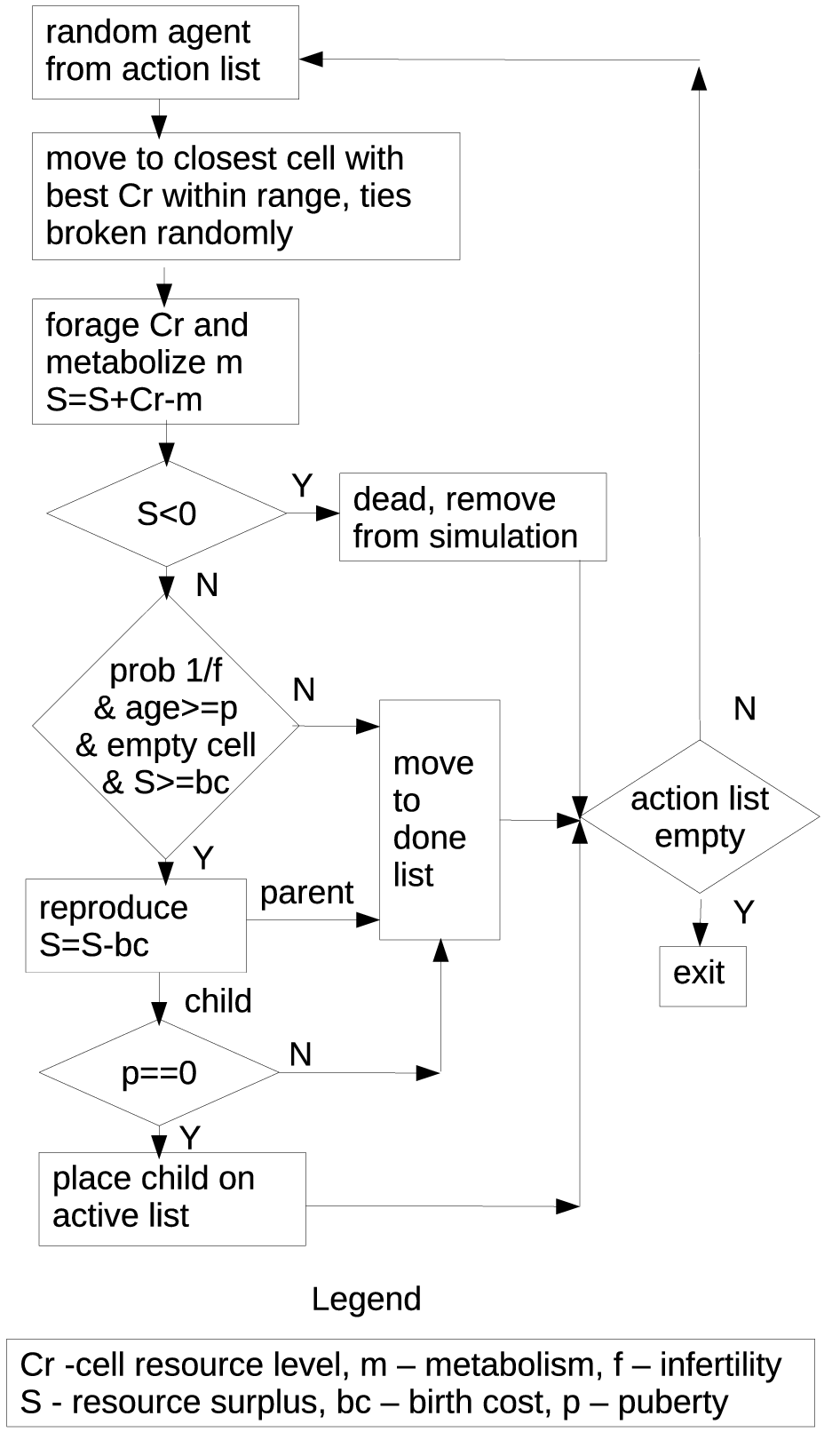}
	\end{center}
	\caption{\small Action cycle for ABM}
	\label{fig:action}
\end{figure}

The paper proceeds by first providing a brief overview of MCT and a detailed discussion of the ABM used. Competitive exclusion and coexistence is then demonstrated for two foraging species with different intrinsic growth rates \citep{milles} in a finite landscape by using an ABM with both resource and space competition. The model is applied to stable and fluctuating environments (seasons) and to single and dual species populations with individual movement. Two separate approaches are used for winter. Detailed population trajectories are presented and analyzed. The regions of coexistence as functions of season length and type, and intrinsic growth are discussed.
% These results are discussed in context with current MCT and community assembly. The stochastic nature of these artificial ecosystems and the non-equilibrium state of not only the population levels but the underlying excess resources provide insight into the density and finite population invasive criterion \citep{chessonEllner,fisher1923,wright1931,cole}. While candidates for competitive coexistence and even commensalism are identified and the artificial ecosystem parameter space is thoroughly explored, the strict invasive criteria of MCT are not met, and quantitative comparisons of the experimental results to spatiotemporal MCT \citep{chesson1981models,chesson2005scale} are beyond the scope of this paper. 
The different perspective that individual, movement-based, discrete stochastic models of a finite population provide relative to MCT theories using mean fields with scaled transitions is highlighted.
% action cycle

\section{Competitive Exclusion Theory}
\cite{hutch} argued \textquote{that diversity of the phytoplankton was explicable primarily by a permanent failure to achieve equilibrium}. If the timescale of environmental change $t_{e}$ is of the same order as the time to competitive exclusion under constant conditions $t_{c}$ (assumed in this case to be summer), then equilibrium cannot be achieved for seasonal ecosystems, and coexistence is then possible. Distributions of both these timescales are determined for two model artificial ecosystems. 

More formally, populations densities $N_{i}$ are given by the Lotka-Volterra model \citep{lotka,volt,kot,chesson2018,chessMech} for two species as: 
\begin{equation}
	\frac{dN_{i}(t)}{dt}=r_{i}N_{i}(1-\sum_{j=1}^{2}\alpha_{ij}N_{i}) , i=1,2
\end{equation}
where $r_{i}$ and $1/\alpha_{ii}$ are the intrinsic growth rate and carry capacity of species $i$, respectively, and $\alpha_{ij} , i \neq j$ is a competitive (commensal) coefficient representing the strength of the effect of species $i$ on species's $j$ intrinsic growth rate. With a single species $i$ population, $N_{j}=0$ and Equation (1) reduces to the Verhulst's continuous and discrete logistic growth equations \citep{verh,murray}. The intrinsic growth rate $r_{i}$ and carry capacity $K_{cc} = 1/\alpha_{ii}$ can be estimated with these equations for the stable and seasonal environments.

For seasonally varying populations, the average over multiple cycles of the seasons is applied to estimate the parameters in Equation (1). For the two-species case, for species $j$ to exclude species $i$ the inequality
\begin{equation}
	\alpha_{ij}> \alpha_{jj}
\end{equation}
must hold \citep{chesson2018}. The mutually invasive criterion for coexistence for two species is based on setting each species resident population at its carry capacity while the other species invades with a population set to small densities \citep{chesson2018,armstrong,kan}. This definition requires that both species as invaders are capable of showing positive growth in these circumstances. That is:
\begin{equation}
	\alpha_{11}> \alpha_{21}
\end{equation}
and
\begin{equation}
	\alpha_{22}> \alpha_{12} 
\end{equation}

For two species to coexist when competing for one resource in a constant environment, two different strategies or niches must be exploited. In the following experiments, the two competing species will have significantly different intrinsic growth rates $r_{i}$. In time varying environments where population equilibrium is not achieved, coexistence may be possible \citep{chessMech,hutch,kan}. Applying this framework to the small, stochastically-driven finite populations of artificial ecosystems requires recognition that stochastic \textquote{survival and reproduction processes .. have finite probabilities of failure} \citep{cole}. Differently seeded runs of an identically configured artificial ecosystem generate a full spectrum of results, from extinctions and exclusions within one season to coexistence for longer than a million generations. The criteria of Hutchinson's time equilibrium and the growth relationships of the $\alpha$ coefficients in Equation 1, however, provide an excellent framework for interpreting the results of these experiments. 

\section{Agent Based Model}
The artificial foraging ecosystem is based on a version of Epstein and Axtell's spatiotemporal agent based model \citep{axtell}. Individual agents forage resources on a two dimensional toroidal landscape (50 cells x 50 cells) with von Neumann neighbors which has replenishing resources at rate $g$ and maximum cell capacities $R_{max}$ defined only by seasons, identical for every cell (a flat landscape). The landscape starts with all cells at $R_{max}$. The action cycle for agents is given in Figure \ref{fig:action}. The agents are defined with vision (6 cells), movement (6 cells), birth costs (0 resources), puberty (1 generation), metabolism (3 resources per generation) and infertility (variable). They are capable of storing all excess resources they forage. The agents die only when they cannot meet their metabolism requirements at the end of an action cycle, otherwise they are immortal. Both the initial agents and those born during the simulation start with zero resources and must forage at least enough resources for their metabolism during their first cycle..  The agents compete both for resources by foraging, and for space for reproduction. The metabolism may be, depending on the season, less than $R_{max}$, allowing surplus resources to be foraged and stored. The species are differentiated by infertility rates. Given the probability $p$ that the agent will actually reproduce, assuming all other requirements for reproduction are met, infertility is defined as $f == 1/p$ and written as $f_{p}$. 

The model simulates both stable and seasonal environments. Winter and summer seasons of equal length are introduced into the model using two distinct approaches as detailed in Table \ref{table:land}. The first approach to modeling winter is to Reduce the Maximum Landscape Capacity (RMLC) for resources in each cell, reflecting less food availability or more difficult foraging. The replenishment rate in this winter model is unchanged from summer and the $R_{max}$ that can be foraged in one generation $t$ is less then the metabolism of the species (identical for all the species). The second approach to modeling winter is to Reduce the Replenishment Rate (RRR) by a given factor $g_{R}$, simulating a slower growth rate of resources in the winter, while allowing the landscape's $R_{max}$ to remain unchanged from summer. This reduction is implemented by limiting the number of cells eligible for regrowth to $C_{total}/g_{R}$ where $C_{total}$ is the total number of cells on the landscape.
%table
\begin{table}[h!]
	\begin{center}
		\begin{tabular}{|c|c|c|c|c|}
			\hline
			Variable & units & Summer & RRR & RLMC \\
			\hline
			$g$ & $r/t$ & 1 & $1/22$ & 1 \\
			$R_{max}$ & $r$ & 4 & 4 & 2 \\
			$K_{cc}$ & agents & 850 & 23 & 0 \\
			\hline
		\end{tabular}
		\caption{Single Season Landscape Parameters where $r$ is resources, $t$ is generations, $g$ is resource regrowth rate, $R_{max}$ is the maximum resource capacity for a cell, and $K_{cc}$ is the single season carry capacity. }
		\label{table:land}
	\end{center}
\end{table}

Random processes are an essential part of generative modeling and artificial ecosystems. The following random processes are at play in these simulations. Initial agent positions are randomly distributed throughout the landscape, though simulations were also run with a constant initial position without significant effect on the distribution of results. As defined above, successful reproduction depends on probabilities of infertility. In each generation, the order of agents selected for action is random. Ties are broken randomly in the agent's movement algorithm. For the RRR winter, random cells are selected each action cycle for replenishment.

\begin{figure}
	\begin{center}
		\resizebox{\columnwidth}{!}{	
			\includegraphics[angle=-90,scale=1.0]{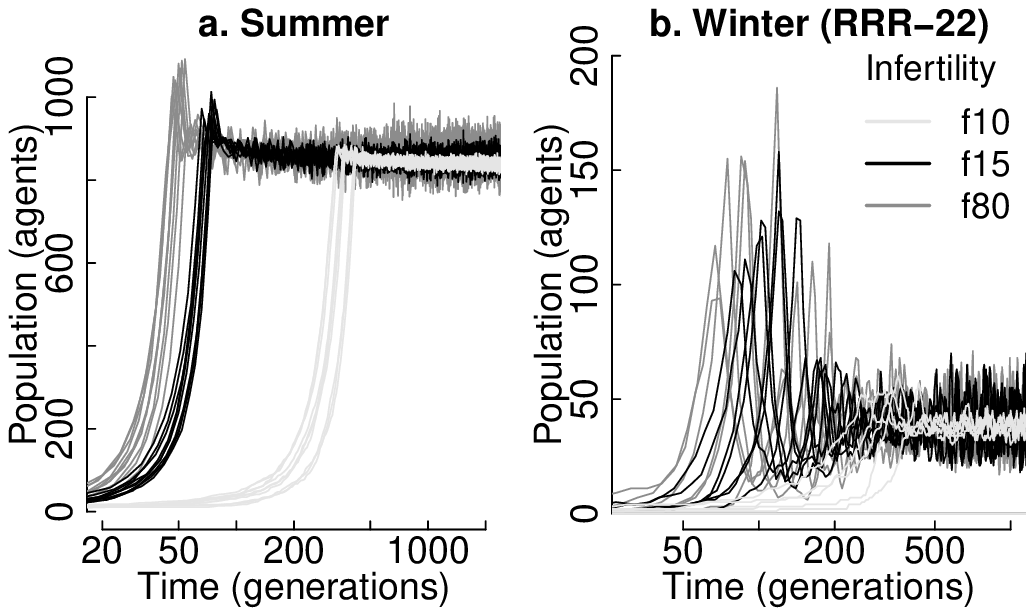}
		}
	\end{center}
	\caption{\small Single species populations in stable environments.} {\small Ten differently seeded runs for $f_{10}$, $f_{15}$, and $f_{80}$ in both summer and winter (RRR-22) seasons.}
	\label{fig:singleRRR}
\end{figure}

\section{Stable Environment}
In the stable environment, which can be either constant summer or winter, each single species' intrinsic growth rate, carry capacity, and equilibrium population are established. The high infertility species ($f_{80}$) is then faced with an invasion of one of two lower infertility species ($f_{10}$ or $f_{15}$) and the resulting population dynamics and extinction times are generated and discussed. The ability to increase population as an invader is part of the invasion criterion for coexistence given in Equation (2). 

\subsection{Single Species Characteristics in a Stable Environment}

The intrinsic growth rate and long term stability for each of these three species during the summer are on display in Figure \ref{fig:singleRRR}a. Ten differently seeded runs for each species sample the numerous random processes discussed above. The initial population is set to ten agents. 
%figure 2
\begin{figure}
	\begin{center}
		\resizebox{\columnwidth}{!}{	
			\includegraphics[angle=-90,scale=1.0]{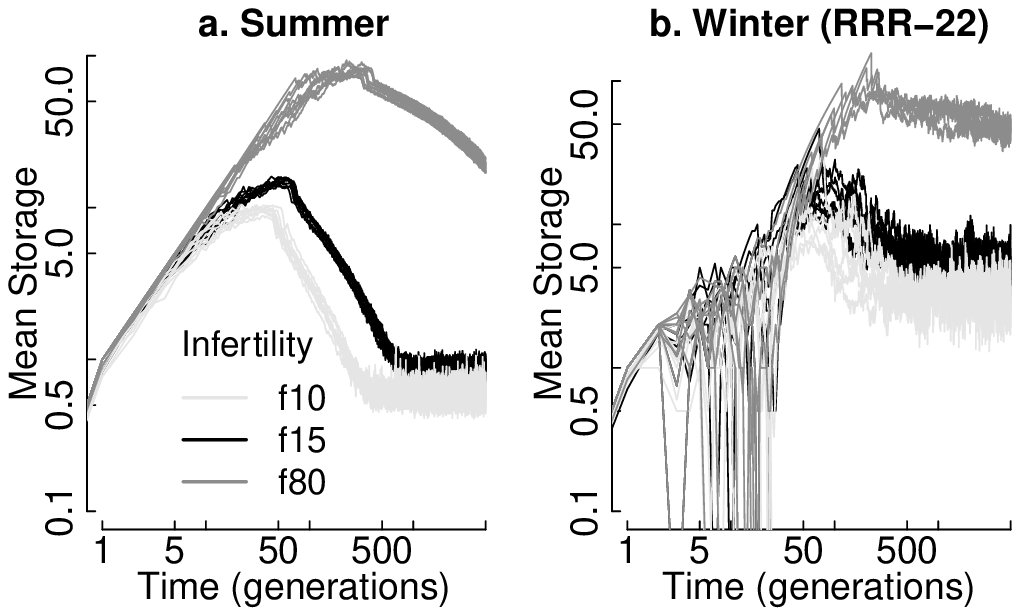}
		}
	\end{center}
	\caption{\small Single species storage in stable environments.} {\small Ten differently seeded runs for $f_{10}$, $f_{15}$, and $f_{80}$ for both summer and winter (RRR-22) seasons. A number of extinctions are evident for the winter season. }
	\label{fig:RRRstorage}
\end{figure}

In Figure \ref{fig:singleRRR}b, the same parameters are presented for the three species during a winter with a 22-fold reduction in resource replenishment (RRR-22). The carry capacity and intrinsic growth rate for all species are, as expected, greatly reduced, and a significant number of runs experience immediate extinction due to the stochastic effects of either long dwell times at low population levels (high infertility: $f_{80}$ loses 50\%) or high population level volatility (low infertility: $f_{10}$ loses 40\%, $f_{15}$ loses 30\%).

These population trajectories demonstrate the model's fidelity to standard discrete Verhulst and Hutchinson-Wright logistic growth equations of mathematical biology and ecology \citep{verh,hutch1,wright,kot,stevenson}.

Figure \ref{fig:RRRstorage} presents the mean stores per individual for the same constant environment ecosystems. The ability to store excess resources is significantly different for different species even in winter. The initial run up in mean storage is due to the low intraspecies competition as the population grows from small initial numbers. Once the population has reached the carry capacity, the mean storage drops down to a steady-state value due to intraspecies competition though the relaxation time to the steady-state mean storage is quite longer than the time to reach carry capacity, significantly extending the period of non-equilibrium dynamics. In fact, equilibrium in mean storage for species $f_{80}$ is not achieved even after 5,000 generations. With a meager mean storage in the height of summer for the low infertility species, population levels are volatile and life is short.
%figure 3 
\subsection{Two Species Competition in a Stable Environment}
\begin{figure}
	\begin{center}
		\resizebox{\columnwidth}{!}{	
		\includegraphics[angle=-90,scale=1.0]{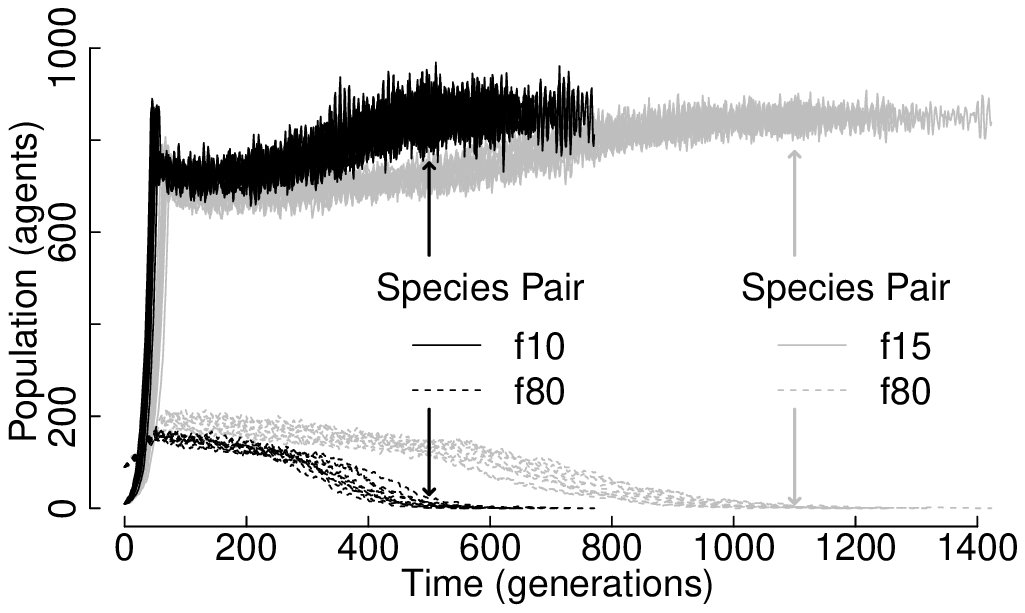}
	}
	\end{center}
	\caption{\small Two Species Exclusion in a Stable Environment} {\small Ten differently seeded runs each of invasive species $f_{10}$ and $f_{15}$ against the resident species $f_{80}$. }
	\label{fig:stableX}
\end{figure}

Figure \ref{fig:stableX} provides the population trajectories for a lower infertility species (initial population 10 agents) invading a higher infertility resident population (initial population 100 agents) in a stable environment. The invader quickly and consistently drives the residents to extinction, demonstrating interspecies selection pressure driving these two pairs of species in a stable environment.

For the resident $f_{80}$ invader $f_{10}$ pair, the resident went extinct at a mean time of $654\pm70$ generations over ten differently seeded runs (maximum time 771 generations). For the same resident pairing with the $f_{15}$ invader, the mean time to extinction was $1223\pm104$ generations over ten differently seeded runs (maximum time 1423 generations). 

These results demonstrate the model's fidelity to the standard Wright-Fisher class, discrete, stochastic, gene-frequency models of mathematical population genetics with selection pressure and overlapping generations \citep{ewens,moran,cannings,stevenson}.

\section{Seasonal Environment}

A periodic variation in environment can lead to opportunities of coexistence which do not exist in constant environments \citep{hutch,chessMech,kan}. Seasons are split into a summer and winter of equal length, and two different models of winter are simulated. The agents will augment their survival with any surplus they may have gained over the previous summer. In the first winter RMLC model, the maximum resources available in a given landscape location for foraging is cut in half, under the agents' daily metabolic requirement. In the winter RRR model, the replenishment rate is reduced by a factor $g_{R}$ which then randomly distributes the resource to the landscape. These single-season models show either reduced or no landscape carry capacity. 

\subsection{Single Species Characteristics in a Seasonal Environment}

\begin{figure}
	\begin{center}
		\resizebox{\columnwidth}{!}{	
		\includegraphics[angle=-90,scale=1.0]{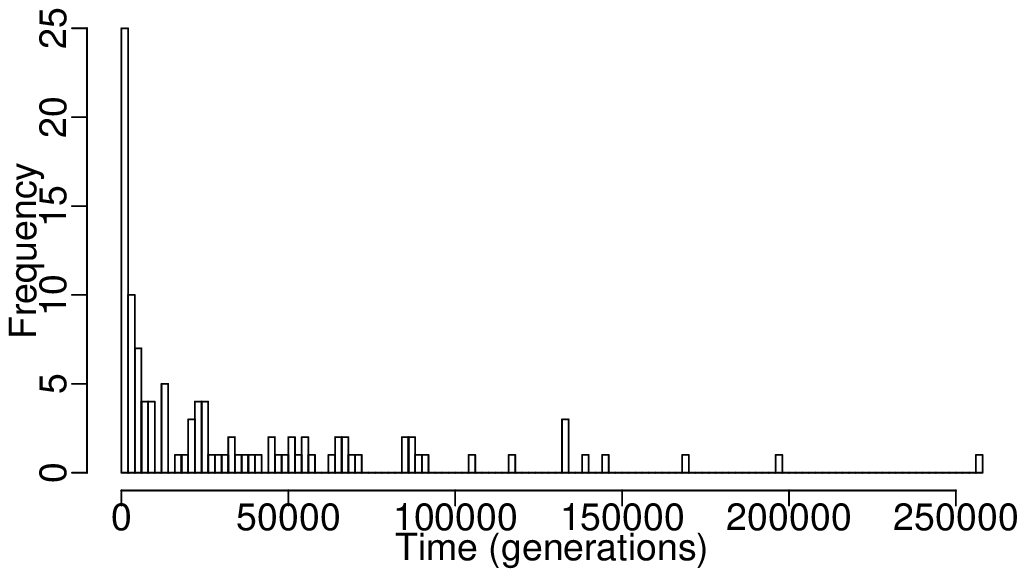}
	}
	\end{center}
	\caption{\small Seasonal Single Species Extinction Times} {\small One hundred ten  differently seeded runs of RMLC seasonal model with species $f_{10}$.}
	\label{fig:seasonSingleXtime}
\end{figure}

For a single species in the seasonal foraging model, the high infertility species $f_{80}$ achieves a stable population in both seasonal models for a very long time (over one million generations) as does the low infertility species $f_{15}$ in the RRR-22 seasonal model. The low infertility species $f_{10}$ in the RMLC seasonal model, however, went extinct within 260K generations with a mean extinction time of 35K$\pm$47K generations for all 110 differently seeded runs. These extinctions are driven by high growth (low infertility) increasing intraspecies competition thus preventing storage of sufficient resources to survive the upcoming winter. Figure \ref{fig:seasonSingleXtime} provides the histogram of these extinction times. In contrast to these extinctions, a commensal effect for this species when invading the $f_{80}$ high infertility species will be shown.

\begin{figure}
	\begin{center}
		\resizebox{\columnwidth}{!}{	
		\includegraphics[angle=-90,scale=1.0]{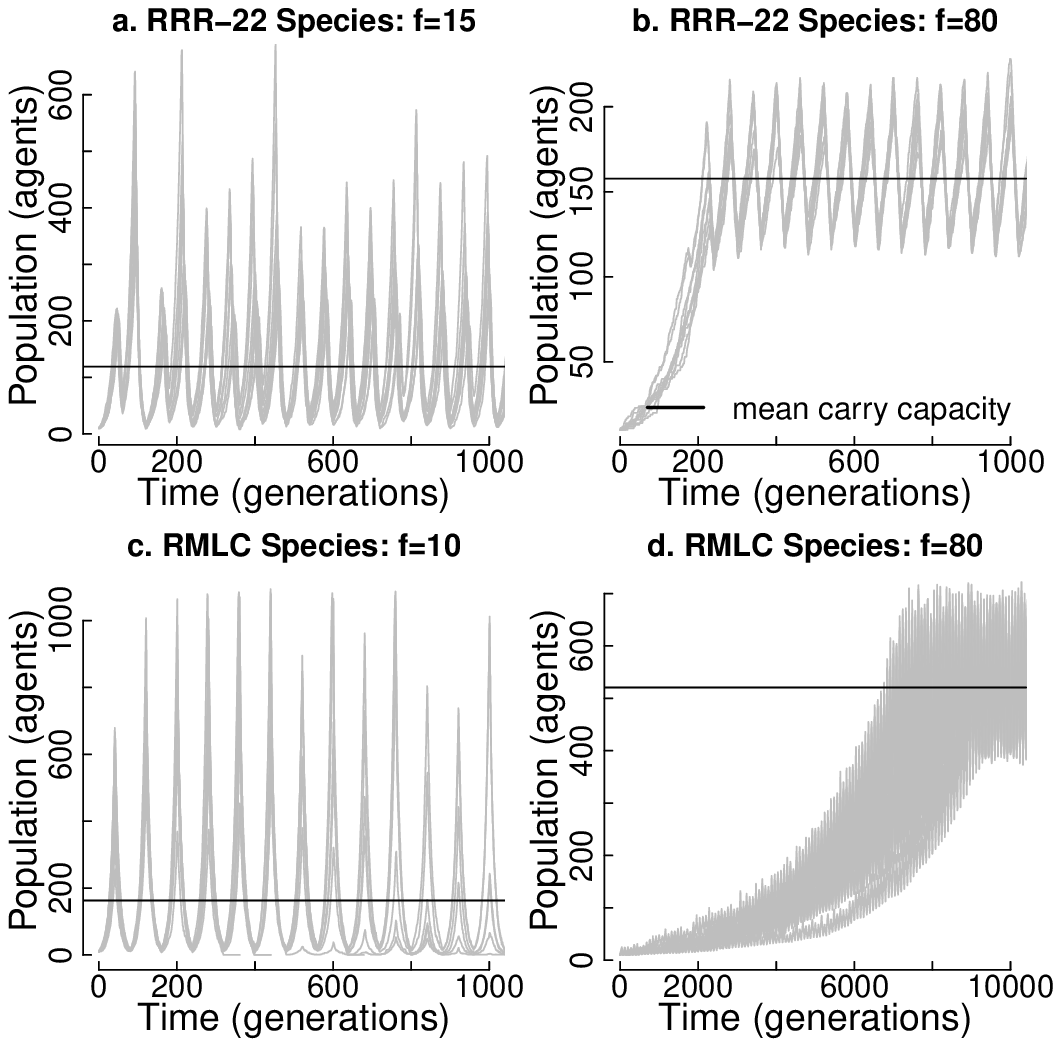}
	}
	\end{center}
	\caption{\small Single Species Population Trajectories} {\small Ten differently seeded trajectories with mean carry capacity: a.) Seasonal RRR-22 with season length 30, species $f_{10}$. b.) Seasonal RRR-22 with season length 30, species $f_{80}$. c.) Seasonal RMCL with season length 40, species $f_{10}$. Extinctions are visible. d.) Seasonal RRR with season length 40, species $f_{80}$.}
	\label{fig:seasonSinglePop}
\end{figure}

For these four conditions, a low and a high infertility in each of the two winter models scenarios, Figure \ref{fig:seasonSinglePop} provides population trajectories and mean carry capacities for each single species. The carry capacities are computed as time-averaged over the population trajectory once the initial growth phase has completed.

The carry capacities $K_{cc}$ in the RRR-22 seasonal model with $g_{R}=22$ reduction in resource replenishment and a season length of 30 generations for selected low and high infertility species ($f_{15}$ and $f_{80}$) are $119\pm94$ and $158\pm25$ agents respectively. For the RMLC winter model with a season length of 40 generations the $K_{cc}$ for selected low and high infertility species ($f_{10}$ and $f_{80}$) are $163\pm196$ and $521\pm77$ agents respectively. These significant differences in average carry capacity support coexistence as the high infertility species carries significant reserves into winter wheres the low infertility species can recover quickly from very low numbers at the end of winter.

The low infertility species are so avaricious that they cannot survive the first winter if started at the resident starting population (100 agents), which is even below the single species seasonal carry capacity expected by the mutual invasive criterion. The higher initial population generates such fierce intraspecies competition that insufficient storage of resources coupled with a rapid degradation of the environment drives immediate extinction. Only by starting these two species at a much lower population (10 agents) can these low infertility species survive the first winter. The high infertility species has no problem surviving the winters with any starting population. This characteristic of the low infertility species means the strict requirement of mutually invasive criterion cannot be met.

\begin{figure}
	\begin{center}
		\resizebox{\columnwidth}{!}{	
		\includegraphics[angle=-90,scale=1.0]{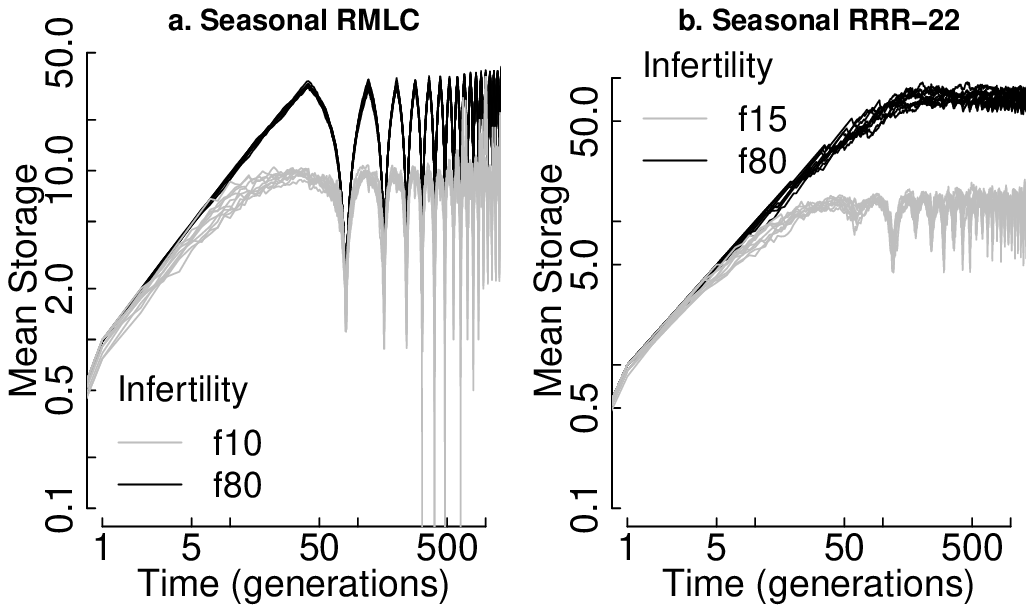}
	}
	\end{center}
	\caption{\small Single species storage in seasonal environments.} {\small Ten differently seeded runs for $f_{10}$ and $f_{80}$ species in the seasonal RMLC ecosystem and for $f_{15}$ and $f_{80}$ species in the seasonal RRR-22 ecosystem. Significant extinctions are again evident for the $f_{10}$ species. }
	\label{fig:seasonSingleStorage}
\end{figure}

The mean storage trajectories for the seasonal single species ecosystems are given in Figure \ref{fig:seasonSingleStorage}. In contrast to the constant environment single species storage trajectories, these trajectories achieve steady state values after just a few seasonal changes. All the species' mean storage, except $f_{80}$ in the RRR-22 scenario, oscillate strongly at the season frequency while the low infertility species do so at much lower levels of mean storage.

\subsection{Two Species Competition in a Seasonal Environment}
 A current metric for coexistence is a mutual invasion criterion which requires that each of the two species be able to grow from low density as an invader into a resident population of the other species \citep{chesson2018,kan}. From Equation 1, $N_{i}$ is small relative to $N_{j}$, which is near the (mean) carry capacity. Equations 3 and 4 then provide the theoretical requirements for stable coexistence. As noted in the previous section, the low infertility species at their seasonal single species carry capacities do not survive the winter. Thus only high infertility residents with low infertility invaders are surveyed for coexistence.

\subsection{Survey of Seasonal Ecosystem Parameters}

The environmental and species characteristics required for competitive coexistence in these two artificial ecosystems are found in a small zone within the larger conditions available to the simulation. For RMLC, season length and  invader infertility are varied while all the other parameters of this ecosystem are held constant. For RRR, the replenishment reduction rate $g_{r}$ and season length are varied.  An exploration of the sensitivity of coexistence to the infertility, season length, and $g_{R}$ parameters provides insight into the conditions necessary for coexistence and increases confidence in the veracity of the simulation. 

\begin{figure}
	\begin{center}
		\resizebox{\columnwidth}{!}{
			\includegraphics[angle=-90,scale=1.0]{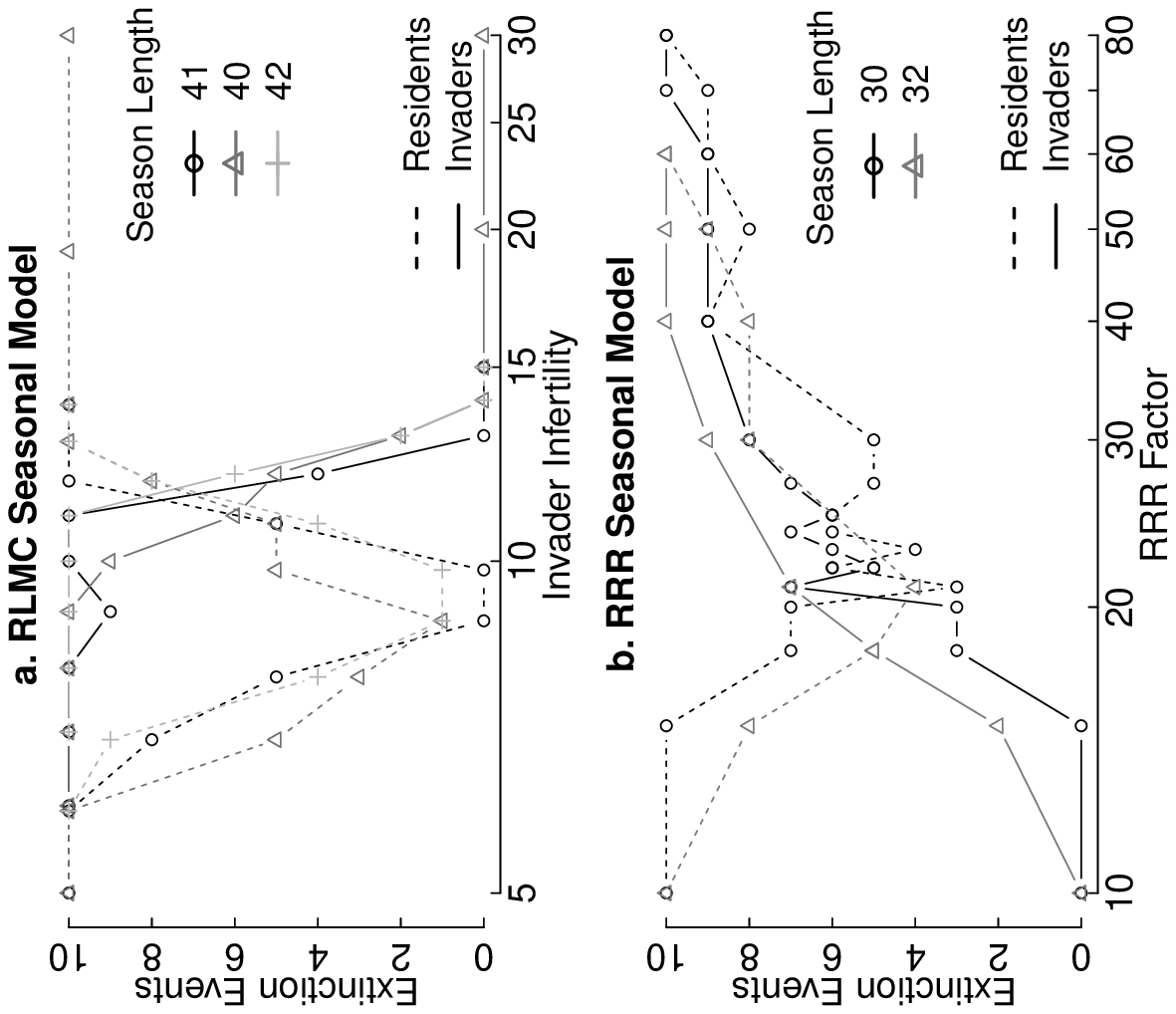}
		}
	\end{center}
	\caption{\small Regions and sensitivities of coexistence zones} {\small For ten differently seeded runs, the vertical axis records the number of invader and resident extinctions for each configuration. a.) Regions of invader infertility that promote coexistence . b.) Regions of winter replenishment reductions that promote coexistence.}
	\label{fig:coXzones}
\end{figure}

For the RMLC seasonal ecosystem, Figure \ref{fig:coXzones}a demonstrates both how the potential zone of coexistence is a function of invader infertility and season length. Starting from the left of the figure, at first the fecundity of the invader is so great that the environment is degraded and neither species survives the first winter. Then, as the invader's infertility increases, the resident begins to show the possibility of surviving the winters. A bit further on, the invader also begins to survive the winters and the possibility of coexistence emerges. As the infertility of the invader continues to increase, the invader begins to outcompete the resident. The resident goes extinct, and the possibility of coexistence disappears. 
% regions 2

For the RRR-22 seasonal ecosystem (with invader infertility $f_{15}$), Figure \ref{fig:coXzones}b provides the sensitivity of the coexistence zone to the replenishment rate of foraged resources. Again beginning on the left, the replenishment rate is sufficient to allow the invader to outcompete and exclude the resident (as it does so powerfully in the summer, Figure \ref{fig:stableX}). As the landscape becomes more resource poor, the resident begins to compete successfully against the invader and the possibility of coexistence emerges. As the landscape replenishment continues to degrade, both species can no longer survive the increasingly harsh winters.

%**** phase diagrams  ****

\subsection{Model Ecosystems}
For each of the two seasonal scenarios a model ecosystem is selected to highlight the various species' population trajectories under a particular ecosystem. For the RMLC climate, a season length of 41 generations and infertility of 10  and 12 were selected. These infertility scenarios, as seen in Figure \ref{fig:coXzones}a, are in the heart of the coexistence zone with a mean absorption time of $310,000\pm334,000$ generations over ten differently seeded runs. In fact, one of the runs still had both species present after 1 million generations. For the RRR model winter, a growth reduction  $g_{R}=21$  was selected with $f_{15}$ and a season length of 30 generations. Figure \ref{fig:coXzones}b shows that this reduced growth rate is also in the heart of the coexistence zone with a mean absorption time of $52,300 \pm 65,500$ generations. This specific configuration also provides examples of both invader and resident extinctions.

The following sections will highlight the stochastic nature of these simulations, which drives large differences in the net results. These two model ecosystems had all three possible outcomes occur due only to different seeds under otherwise identical conditions: exclusion occurred for both species, intermediate length coexistence occurred with either species eventually going extinct, and coexistence occurred for long to very long times.

\subsection{Reduced Maximum Landscape Capacity Seasonal Model Ecosystem}

\begin{figure}
	\begin{center}
	\resizebox{\columnwidth}{!}{
		\includegraphics[angle=-90,scale=1.0]{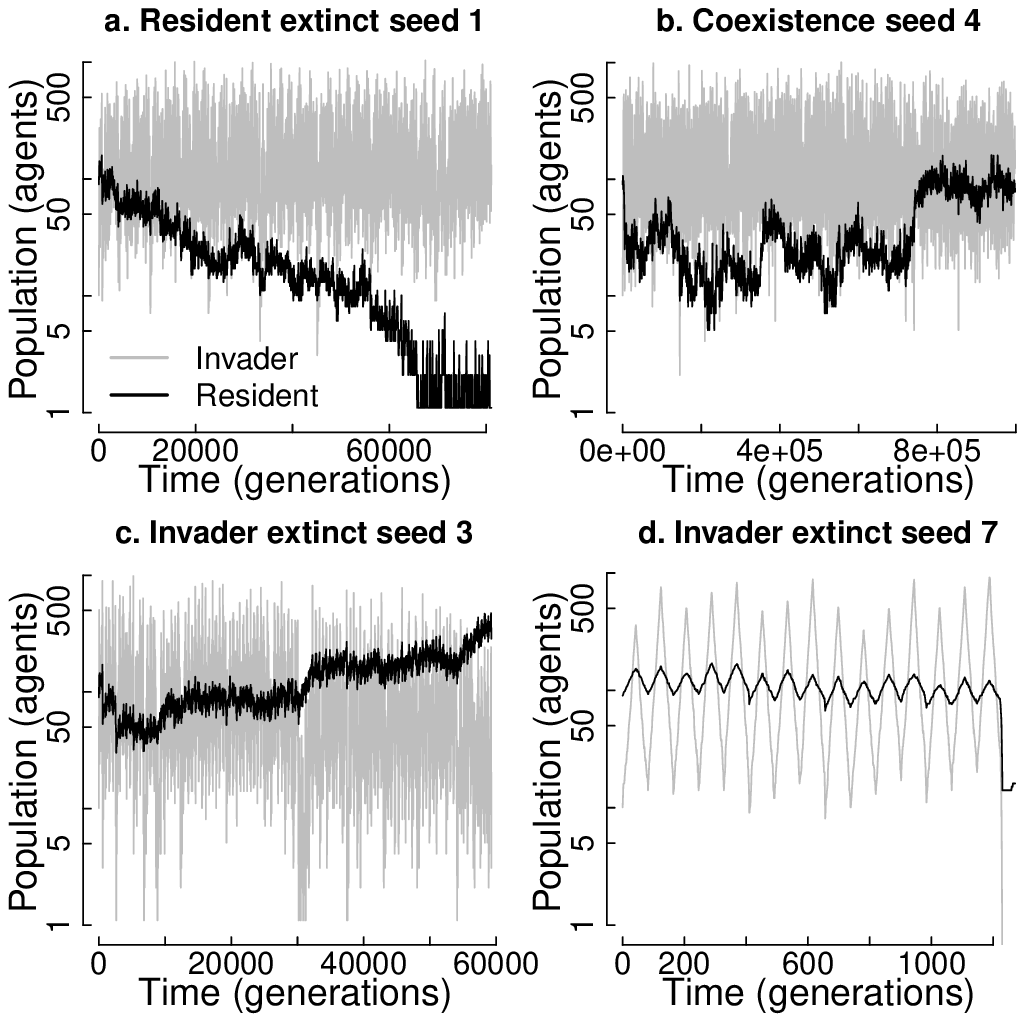}
	}
	\end{center}
	\caption{\small Trajectories for Seasonal RMLC model ecosystem} {\small All four possible outcomes from ten runs with different seeds on an otherwise identical seasonal RMLC model ecosystem. Infertility $f_{12}$ for Figures a and b and  $f_{10}$ for Figures c and d.}
	\label{fig:rmlcCox}
\end{figure}

Figure \ref{fig:rmlcCox} provides species population trajectories for samples of four differently seeded simulations of the model RMLC ecosystem. Two invader extinctions are shown. Figure \ref{fig:rmlcCox}d is a relatively immediate extinction at 1200 generations which reveals both the oscillatory effect of the seasons on species' populations and a rather harsh winter for both species resulting in the invader's extinction. The relative harshness of differently seeded winters is not due to any environment differences but rather accumulated effects from stochastic spatial competition, reproduction, and foraging. Figure \ref{fig:rmlcCox}c demonstrates an overall positive population growth for the resident species which results in the invader's extinction after 60,000 generations. Figure \ref{fig:rmlcCox}b provides a strong example of coexistence with both species thriving for over 1 million generations (surely a \textquote{very long time}). Since the invader species is highly unlikely to ever survive by itself (as shown in Figure \ref{fig:seasonSingleXtime}) and the resident easily survives, the effect is commensal. In these highly volatile population scenarios, the agents move in spatial waves, compressed into tight wavefronts with the landscape either barren behind or just fully restored ahead. \citep{stevensonEcon}. Such spatial compression reduces the fecundity of the invader. Without the space occupied by the higher infertility resident species sharing the wavefront, the invader would go extinct. Finally, Figure \ref{fig:rmlcCox}a shows a steadily declining resident population with resultant extinction after 80,000 generations. For Figures \ref{fig:rmlcCox}b and \ref{fig:rmlcCox}c, the trajectories argue strongly for coexistence over \textquote{a very long time}.

\subsection{Reduced Replenishment Rate Seasonal Model Ecosystem}

\begin{figure}
	\begin{center}
		\resizebox{\columnwidth}{!}{
		\includegraphics[angle=-90,scale=1.0]{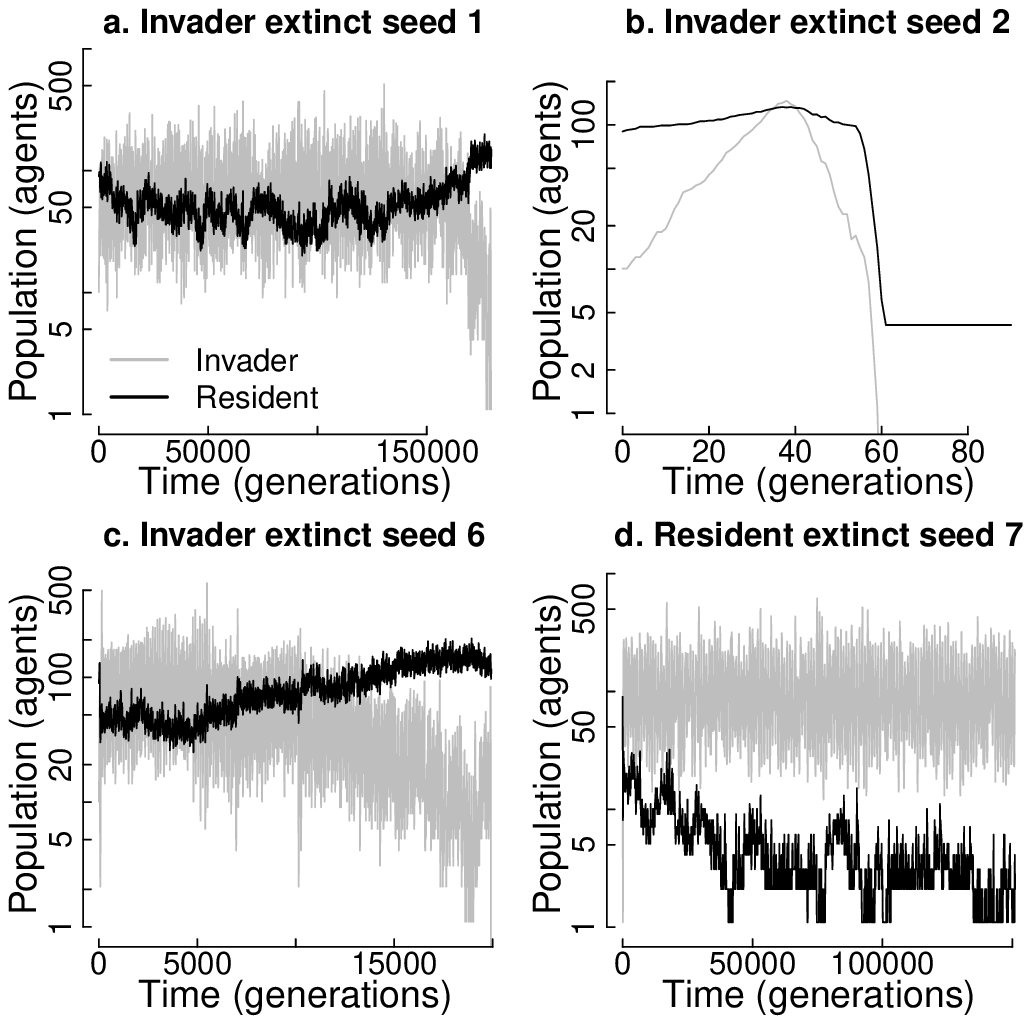}
	}
	\end{center}
	\caption{\small Trajectories for Seasonal RRR-21 model ecosystem} {\small All four possible outcomes from ten runs with different seeds on an otherwise identical seasonal RRR-21 model ecosystem.}
	\label{fig:RRRcoX}
\end{figure}

Rather than reducing the foraging efficiency, the model RRR-21 ecosystem simulates a winter where the landscape is unable to replenish the resources as quickly as it does in the summer. Figure \ref{fig:RRRcoX} provides species population trajectories for a sample of four differently seeded simulations of a model ecosystem with this RRR-21 winter. Figure \ref{fig:RRRcoX}b records an immediate extinction in the very first winter of the invader species. Though only a few resident agents remain, they will most likely propagate to the carry capacity as exemplified by Figure \ref{fig:seasonSinglePop}b. A long drawn out extinction of the resident species after 150,000 generations is shown in Figure \ref{fig:RRRcoX}d, exemplifying the difficulty of eliminating a small fraction of a large stochastic population \citep{ewens}. Note the resident population was down to one agent numerous times, the first around generation 50,000. And yet exclusion was more than 100,000 generations away. Figure \ref{fig:RRRcoX}c shows a slow positive trend for the resident population resulting in an invader's extinction after 20,000 generations, a possible candidate for coexistence. Figure \ref{fig:RRRcoX}a records a strong candidate for coexistence with constant populations of both species for over 130,000 generations with a increasing resident population after generation 130,000, leading to an invader extinction at 180,000 generations.

\section{Discussion}

%An agent based model of a simple foraging ecosystem is shown to generate complex behaviors associated with two species competition for one replenishing resource. These species differ only by intrinsic rate of growth. The lower infertility species strictly excludes the higher infertility species in a constant (summer) environment ecosystem. Though the higher infertility species, reproducing at a slower rate, is able to harvest greater stores of resources before the carry capacity is reached, the pressure of natural selection in a constant environment sees no long term advantage for this initial, larger storage. 

%hutch non-equi
%exclude in consant coex in seasonal only
%coxeistence as commensal
%origin restrictions

The complex, non-equilibrium behaviors of a finite population of two species making individual movement decisions and competing for the same space and resources in constant and seasonally varying environments were successfully modeled with an underlying mechanism independent from those normally used for theoretical and computational modeling of persistence and extinction \citep{hening2021general}, demonstrating the utility of an artificial life approach. Only with seasonally varying resources would coexistence emerge, it's emergence could not be guaranteed for every randomly seeded run, and commensalism also emerged. The dynamics of finite-time coexistence in a stochastic, finite population were addressed, and limits on the possible origins of two-species, coexistent ecosystems were delineated.

\cite{hutch} argues that coexistence is possible if a \textquote{permanent failure to achieve equilibrium} occurs. The single species seasonal mean storage trajectories (Figure \ref{fig:seasonSingleStorage}), the population trajectories (Figure \ref{fig:seasonSinglePop}), and the mean times to extinction for constant environment (Figure \ref{fig:seasonSingleXtime}) all demonstrate that equilibrium will not be achieved within seasonal cycles $t_{e}$ of 30 to 42 generations, thus supporting the possibility of coexistence while also emphasizing the need for non-equilibrium modeling.

In the seasonal ecosystem, zones of potential coexistence are identified based on the infertility of two competing species, the length of the seasons, and the modeled winters. Due to the large number of stochastic spatial and reproductive decisions made in the generation of this artificial ecosystem, any one run in the identified zones of potential coexistence can result in fast extinction of either or both species, a slow trend to extinction for either species, coexistence for \textquote{a very long time}, or even commensalism. While MCT describes these spatial and temporal ensembles as mean density fields with scale transition factors, the actual population trajectories provide a valuable perspective for what might be observed in laboratory field work.

Varying resource environments modeled on seasons reveals the fragility of low infertility species. One species ($f_{10}$) was shown to be highly likely to go extinct in the RMLC seasonal model (Figure \ref{fig:seasonSingleXtime}). But by adding the high infertility species as a resident, this two-species ecosystem improved the survivability of the $f_{10}$ species compared to its single species survivability, demonstrating a commensal effect. The inability of a low infertility species to survive in seasonal single species ecosystems suggests that the origin of a coexistence ecosystem would only occur as an invasion of a low infertility species into a high infertility resident population.

The significant sensitivity of persistence and extinction events to the underlying individual stochastic movement processes for finite populations within the same ecological parameters (species and environment) provide an important perspective on competitive exclusion and coexistence. This artificial ecosystem approach supports study of possible niches for coexistence based on spatial, seasonal, individualistic movements, and metabolic differences; and allows scaling to larger metacommunities.
%Both low infertility models would immediately go extinct in both seasonal models if the starting population was the same order of magnitude as their seasonal single species carry capacity. This behavior prevents a positive indication of competitive coexistence using a strict mutual invasive criterion.

%The artificial ecosystem provided details on the complex relationship between species intrinsic growth rate and storage capability, the environmental carry capacities, and the zones of potential coexistence. While qualitative insight has been gain by using MCT, no direct quantitative mapping of these results to the theory was achieved.

\footnotesize
\bibliographystyle{apalike}
%\bibliography{compX2.bib} % replace by the name of your .bib file
\bibliography{/home/jackcs/biblio/feb03_23.bib} % replace by the name of your .bib file

\end{document}